\documentclass{IEEEcsmag}

\usepackage{graphicx}
\usepackage{subfig}
\usepackage{caption}
\usepackage{comment}
\usepackage{url}
\begin{document}

\sptitle{General submission}

\title{Measuring the Energy of Smartphone Communications in the Edge-Cloud Continuum: Approaches, Challenges, and a Case Study}

\author{Chiara Caiazza}
\affil{Dip. di Ingegneria dell'Informazione, Universit\`a di Pisa, Pisa, Italy}

\author{Valerio Luconi}
\affil{Istituto di Informatica e Telematica, Consiglio Nazionale delle Ricerche, Pisa, Italy}

\author{Alessio Vecchio}
\affil{Dip. di Ingegneria dell'Informazione, Universit\`a di Pisa, Pisa, Italy}

\begin{abstract}
As computational resources are placed at different points in the edge-cloud continuum, not only the responsiveness on the client side is affected, but also the energy spent during communications. We summarize the main approaches used to estimate the energy consumption of smartphones and the main difficulties that are typically encountered. A case study then shows how such approaches can be put into practice. Results show that the edge is favorable in terms of energy consumption, compared to more remote locations.
\end{abstract}



\maketitle

\section{Introduction and Background}

With edge computing, computational resources are pushed to the periphery of the network, close to the end-users, to achieve reduced latency~\cite{Satyanarayanan17:emergence}. Edge computing, however, should not be considered as a substitute for cloud computing: they can co-exist giving rise to a continuum where the placement of resources can be determined to optimize quality and costs~\cite{Dustdar23:distributed}.
An aspect that has not been explored sufficiently is the impact of reduced communication latency on the energy required to communicate.
Most of the literature concerning edge/cloud computing and energy consumption focuses on saving energy by offloading part of the computation from client devices to in-network computing elements. A method for reducing energy usage based on cooperative task offloading on the edge was presented in \cite{Zahed2020:green}. Research also focused on power consumption and latency in hierarchical edge networks~\cite{Zhang2018:energy}. A computation offloading strategy that is suitable for the Industrial IoT domain is presented in~\cite{Hazra23:cooperative}. Experimental evaluation of the advantages of offloading workloads to edge or cloud servers has been discussed in~\cite{Hu2016:quantifying}. A survey on energy-aware edge computing can be found in~\cite{Jiang2020:energy}. The purpose of these studies is to weigh the benefits and drawbacks of shifting computational work to external servers.

Other works focused on measuring energy consumption due to communications on smartphones. In~\cite{Chen2015:smartphone}, a model of the LTE interface was derived and communications were identified as a major contributor to smartphones' energy consumption. Subsequent work aimed to balance energy and latency requirements by manipulating the LTE interface parameters~\cite{ Brand2020:adaptive}. Experimental approaches were used to compute the energy consumption on smartphones when using specific protocols, e.g. DASH~\cite{Zhang2016:profiling}. Our recent work focused on estimating communication energy consumption via analytical and experimental approaches~\cite{Caiazza2021:edge, Caiazza22:saving}.

In this article, we provide the following contributions:
\begin{itemize}
    \item We discuss different approaches to measuring the energy consumed by smartphones in the edge-cloud continuum, highlighting their benefits. We do not consider computation offloading, as, despite the benefits, it cannot always be applied: identifying computing units to be relocated can be difficult, and systems have to be redesigned and reimplemented.
    \item For each approach, we discuss the main challenges that need to be faced by researchers/experimenters.
    \item We present a case study, to share our experience and the practical solutions adopted to overcome the previously identified challenges. 
    \item The case study shows that communication with edge-placed resources is generally more efficient in terms of energy required. Our results focus on smartphones but apply also to other types of constrained devices.
\end{itemize}

\section{Approaches}

The energy needed by a client for communicating can be estimated through analytical methods, simulations, and software/hardware monitors. 

\emph{Analytical} techniques model the energy expenditure of communication. The most relevant states of the interface are identified, then a deterministic or stochastic model of the traffic generated by an application is used to compute the overall costs. The interface alternates sleeping and awake periods, according to different duty cycles. The energy needed during communication with duration $T$ is equal to 

\begin{equation} 
E_T = P_{sleep}*T_{sleep} + P_{awake}*T_{awake}
\label{eq:1}
\end{equation}

\noindent with $P_{sleep}$  and $P_{awake}$ the power needed in the sleep and awake states, $T_{sleep}$  and $T_{awake}$ the time spent in the sleep and awake states, and $T = T_{awake} + T_{sleep}$. Discontinuous reception (DRX) is a mechanism adopted by LTE/LTE-A to reduce the energy needed when no packets are sent or received in a short time. The $T_{awake}/(T_{awake} + T_{sleep})$ ratio in DRX determines the energy savings, as $P_{sleep}$ is significantly smaller than $P_{awake}$. The model of the interface is then combined with the application model: voice, for example, is characterized by an on/off behavior, with an exponentially distributed duration of talk spurts and silence. Markov and semi-Markov models have been frequently used in analytical studies~\cite{Ramazanali16:performance}. 
The main advantage of analytical techniques is the possibility to obtain a closed form of the metrics of interest. This allows designers to easily evaluate the impact of changes to the parameters of operation, or to explore low probability tails, which are difficult to obtain with other methods.

\emph{Simulations} rely on a power model of the interface that is driven by network events occurring in the considered scenario. Extensions that model the energy consumption and the battery of user terminals are included in popular simulators, such as NS-3~\cite{Pasca16:ns3}. Simulators allow researchers to easily change the network topology and characteristics to explore the impact of different placements in the edge-cloud continuum. 
 
\emph{Software monitors} provide an experimental estimate of energy consumption with no hardware support. Dumpsys\footnote{\url{https://developer.android.com/studio/command-line/dumpsys}} collects information about the usage of resources, including energy, on Android devices. Dumpsys can collect battery diagnostic information such as discharge steps, connection to external power, and on/off screen events. The output of dumpsys includes application- and communication-related events that are significant from the point of view of energy consumption: status and usage of wireless connections, execution of jobs, and possession of wake-locks (mechanisms used by applications to request the OS not to enter deeper sleeping states or to keep the screen on). Dumpsys’ output can be parsed for the events of interest or can be provided to Battery Historian\footnote{\url{https://github.com/google/battery-historian}}, which depicts the battery consumption over time along with communication-related events, network conditions, application execution, and usage of energy-demanding subsystems. Battery Historian also estimates the fraction of energy needed by a specific application. The output of dumpsys can be fed to tools adopted in the context of software engineering, to evaluate the energy wasted by code smells~\cite{DiNucci17:software}. Unfortunately, a tool that evaluates energy waste due to wrong communication patterns does not exist. Software monitors generally rely on power profiles, provided by device vendors, that quantify the additional current needed by system components. It is also possible to generate online a power model of a given subsystem, as the relationship between the energy spent in a period and the utilization rate of a resource: the first can be estimated by observing the state of charge of the battery and the second by means of OS-level counters (e.g. via the \texttt{proc} filesystem)~\cite{Hoque2015:modeling}. Other approaches use the instantaneous variations in the battery voltage to infer the additional current required by a subsystem~\cite{Xu2013:vedge}. More recent devices include a Hardware Abstraction Layer (HAL) that provides power consumption information on the different sub-system rails. The Power Stats HAL is particularly suitable for nonsteady conditions, as it returns instantaneous information collected through an onboard power-monitoring circuit. A trace containing the estimated current flowing on the different rails can also be collected through the Perfetto\footnote{\url{https://perfetto.dev/}} tracing suite. 
Similar mechanisms are available for iOS-based devices, where energy logging can be activated to monitor the energy efficiency of apps. 

\emph{Hardware monitors} can power the device under test and, at the same time, measure its power consumption with great accuracy. Hardware power monitors are generally controlled through a set of APIs to automate tests. Collected traces can be stored for further analysis. These traces can integrate signals received from the device under test via UART. This is useful to avoid synchronization problems due to the different clocks and highlight application-originated events. 
Examples of power monitors used in literature include the Monsoon and Agilent N6705 DC Analyzer.
Differently from software-based monitors, hardware ones do not impose computational load on the device under test even at high sampling rates.

\section{Challenges}

\subsection{Analytical and Simulation Approaches}
Despite the flexibility of analytical and simulation approaches, there still are some challenges. A model that captures all the intricacies of network communication protocols would generate highly complex problems hard to solve from an analytical standpoint, or difficult to simulate in reasonable time. Reducing such problems is mandatory, generally by introducing assumptions or approximations. The consequence is a loss of the model's fidelity and results' accuracy. Analytical or simulation approaches, despite being extremely useful for the early stages of design or evaluation, can be representative of some special cases. Moreover, models and simulators need to be validated against real-world measurements. 

To obtain more realistic results or to reduce the complexity, certain parameters can be extracted from real-world measurements (e.g., the power model of a real LTE interface). Such values can be obtained via existing datasets or previous literature, or direct measurements.

\subsection{Software and Hardware Approaches}
The main issue with software/hardware monitors is that the device under test and the surrounding environment are not completely under control. Background OS tasks can interfere with energy measurements. Moreover, multiple apps can compete for the communication module. Isolating the power consumption of a single application is difficult: hardware power monitors only measure the aggregate consumption; software monitors provide per-application data, but inaccuracies cannot be avoided as some subsystems may enter a different power level because of other apps or system tasks. Considering that the load imposed by system tasks is high at startup, e.g. because of system or software updates, one could avoid running measurements just after the boot phase, or discard the initial part of experiments. To mitigate the effects of unpredictable system tasks, experimenters can adopt data analysis techniques, e.g. averaging the power consumption over a long period or selecting the minimum consumption over multiple slots (under the assumption of a constant communication rate and that uncontrolled tasks can only increase the overall consumption). Other strategies may involve running baseline experiments to obtain the system power consumption when idle.

Challenges also arise from the external environment. In wireless networks, the transmission quality depends on signal strength, which impacts the transmission power and duration, the number of retransmissions, and energy consumption too. To cope with this problem, smartphone OSes provide APIs for obtaining the signal strength, cell identifier, etc. Correlating this information with the consumed energy helps better understand whether an anomalous consumption is due to the communication itself or to external, and potentially confounding, factors.

\subsubsection{Software monitors}

The main issues of software monitors regard determining the accuracy of measures and their need for calibration. A software monitor estimates the energy depleted in a period and attributes it to the responsible applications. The first phase - estimation - relies on the power models of the device subsystems which can be based on utilization rates (e.g. the CPU) or FSMs (e.g. the radio). This can require determining the coefficients linking the dependent variable (the energy) with the independent one (e.g. CPU frequency). The FSM could be derived according to a black-box approach, thus possibly introducing inaccuracies~\cite{Hoque2015:modeling}. The attribution of the energy spent in a period to applications can be inaccurate, for example, due to the limited sampling rate of independent variables or because the status of subsystems generally depends on their \emph{combined} interaction with applications. In some cases, software monitors are unable to assign a fraction of the energy spent to applications, marking it as ``unaccounted''~\cite{source}.
Finally, software monitors can perturb the system under observation, by contributing to battery consumption~\cite{Yoon2012:appscope}.

\subsubsection{Hardware monitors}

Using a hardware power monitor is less immediate, compared to software-only solutions. Hardware power monitors can be accompanied by software tools that ease the collection, visualization, and analysis of traces. However, despite the possibility of automating measurement processes, the execution of repeated experiments is more labor-intensive. 
Connecting a power monitor to a smartphone is challenging. The monitor generally acts as a power supply for the device, thus battery-operated devices require modifications to receive power from an external source. This involves physically disconnecting the battery cell from the rest of the smartphone's circuits and connecting the monitor's power supply to the circuits between the cell and the motherboard. Moreover, hardware power monitors do not allow estimating the energy consumption of devices in the wild or in movement. In some scenarios, the consumption of an application can differ from user to user or when changing the surrounding environment.

\begin{figure*}[!t]
    \centering
    \includegraphics[width=0.65\textwidth]{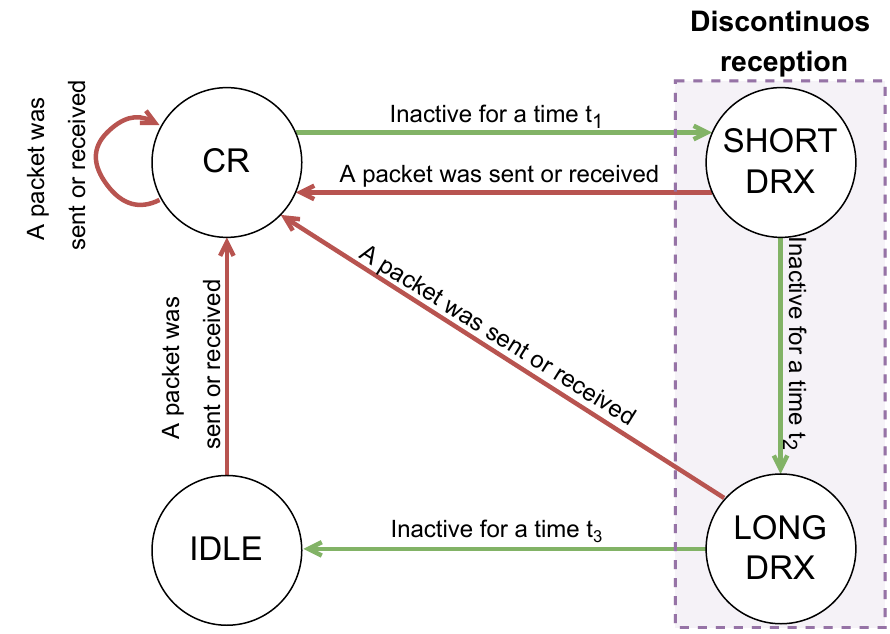}
    \caption{The finite state machine of an LTE module.}
    \label{fig:ltefsm}
\end{figure*}

\section{Case Study: The Impact of Latency in the Edge-Cloud Continuum}

We describe our experience in measuring the energy spent by a client when communicating with servers in the edge-cloud continuum. The setup is composed of a smartphone connected to the network of an Italian MNO. The smartphone communicates with servers placed on the edge, on the cloud, or on a far cloud. The client operates according to a request-response scheme, issuing requests with a fixed frequency. The results are obtained using an analytical approach, a software monitor, and a hardware monitor.

\subsection{Analytical Approach}

We modeled the LTE interface as an FSM with four states: CR (Continuous Reception), SHORT DRX, LONG DRX, and IDLE (Figure~\ref{fig:ltefsm}). The most energy-demanding state is CR, while the least one is IDLE. DRX states are used for gradually transitioning from CR to IDLE. When the interface sends or receives a packet, it enters CR. When the transmission is complete, if there are no other packets, the interface transitions to less energy-demanding states after a fixed interval. The more energy-saving the state, the longer the time interval ($t_1 < t_2 < t_3$). To reduce consumption, in SHORT DRX, LONG DRX, and IDLE, the interface alternates sleep to wake-up periods, used to check for incoming packets. Sleep phases are longer if the state is less energy-demanding (Equation~\ref{eq:1}), making the interface less responsive.
To cope with the intricacies of a transport protocol like TCP, we experimentally collected the transmission and reception times and fed them into the model. 

We computed the ratio between the energy needed to download resources in the 2-12 MB range from an edge and a cloud server. The edge is more efficient, needing 54-40\% of the energy needed by the cloud server (with better results for larger resources). This happens because reduced latency allows reduced transmission and reception times (so shorter times in more demanding states). More details can be found in~\cite{Caiazza2021:edge}.

\subsection{Software Monitor}

We implemented the client as an Android app.
Periodically, the app sends an HTTP request and waits for the response. The payload of the request is varied to understand the effect of the amount of transferred data on energy consumption. The server simply echoes back the payload, as our focus is on the energy required for communication, independently from the application logic.
The setup is composed of a smartphone connected to an LTE operator, and a server hosted by GARR, the Italian research network. To emulate cloud and far cloud servers we artificially added delays to the incoming and outgoing traffic on the server with \textit{tc}, 100 and 200~ms respectively. To reduce interference, we removed and disabled all unnecessary applications from the smartphone. More details can be found in~\cite{Caiazza22:saving}.

\begin{figure*}[t!]
    \centering
    \includegraphics[width=0.48\textwidth]{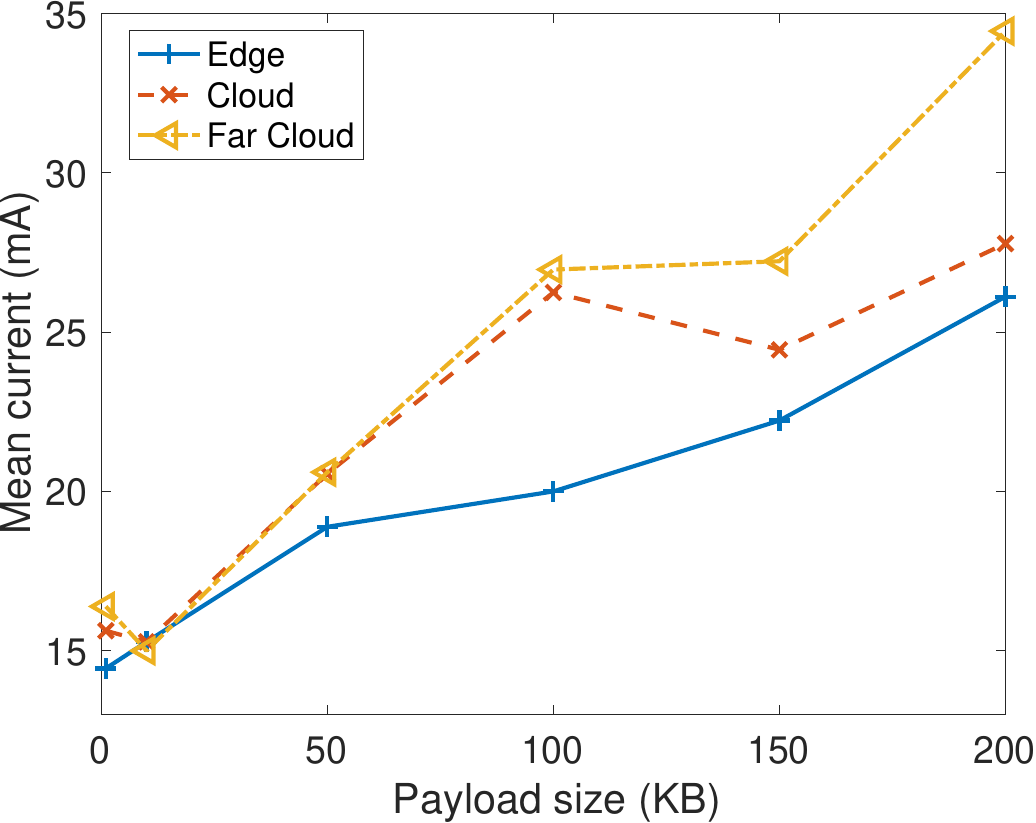}
    \caption {Absorbed current for the three edge/cloud configurations, measured with a software monitor. For each payload size, we selected the minimum value of mean absorbed current computed for slots of one hour each. In this case, the cloud and the far cloud servers are emulated by artificially adding 100 ms and 200 ms to the path with \textit{tc}.}
    \label{fig:softwareMonitor}
\end{figure*}

We ran our application in the three edge/cloud configurations, with payloads varying from 1 to 200 KB. We used dumpsys to collect the battery level during the experiment. The battery of the smartphone was completely charged before every run. To ensure a limited amount of interference from system processes, before the result analysis, we collected the extra amount of traffic produced by the system, which we found to be small. Thus, the disturbance caused by unwanted traffic is limited. To ensure correct operations, we measured the time to completion of each HTTP request, which showed variability during the initial part of the experiment. We thus discarded the first two hours and kept the remaining seven hours, to evaluate only the steady-state operations.
To avoid other confounding factors, we further enhanced the data cleaning. We divided data into one-hour slots and computed the average current absorbed in each slot. We considered the slot with the lowest value of absorbed current as the most accurate since the higher values obtained in other slots could be due to external factors (e.g. bad channel conditions, concurrent tasks).
Figure~\ref{fig:softwareMonitor} shows the results. The most convenient choice in terms of energy consumption is the edge, followed by the cloud, and the far cloud, respectively. The cloud solutions can consume up to 35\% more than the edge one, for certain payload configurations, and show very similar energy consumption for payloads up to 100 KB.

To confirm such results, we ran the app until the complete discharge of the phone. The payloads ranged from 50 to 200 KB. We avoided using smaller payloads as in the previous experiment we noticed that the results for those values were not particularly significant. Being the duration of the experiment rather long ($\sim$10 hours for each sample), we preferred not to run scarcely useful tests. We normalized the device uptime, as when the device reports the battery as fully charged, the actual Coulomb charge can differ. To obtain a fair comparison, we normalized the device uptime by the charge measured at the beginning of the run using dumpsys.
The edge configuration obtained longer device uptimes by 5-23\% compared to the two cloud configurations for payloads of 100-200 KB. In particular, for the 100 KB payload the edge configuration obtained an improvement of 6\% and 9\% with respect to the cloud and far cloud, respectively. For the 150 KB payload, 17\% and 23\%, and for the 200 KB payload, 8\% and 14\%. The rather large variability, considering the high duration of experiments, could be ascribed to the fluctuations of the network conditions, over which we had no control. Note that, in this case, we did not apply any filtering to the collected data. For a 50 KB payload the edge configuration is characterized by an even longer uptime.

\subsection{Hardware Monitor}

\begin{figure*}[t!]
    \centering
    \includegraphics[width=0.5\textwidth]{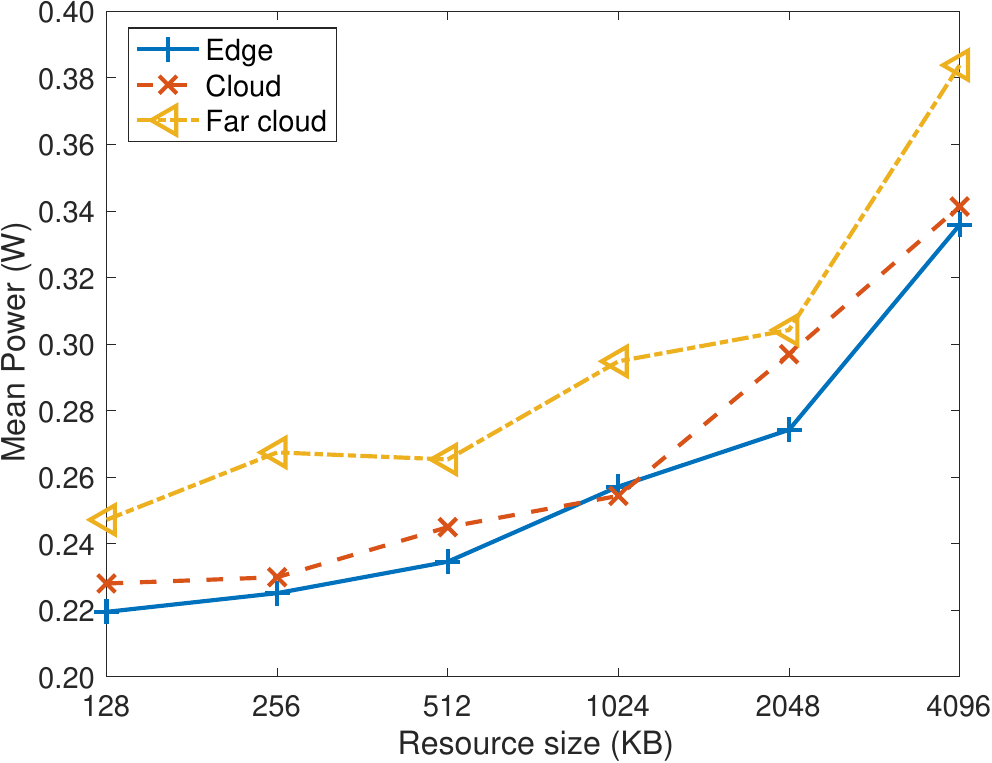}
    \caption {Power consumption for the three edge/cloud configurations, measured with a hardware power monitor. For each resource size, we select the minimum value of mean power computed for slots of 10 minutes each. Differently from the software monitor use case, here, edge, cloud, and far cloud servers are real-world servers from the GCS infrastructure. In particular, they are located in Italy, South Carolina, and Australia.}
    \label{fig:google-HTTP2}
\end{figure*}

We used the Otii Arc\footnote{\url{https://www.qoitech.com/otii-arc-pro/}} power monitor to measure the energy consumption.
The Otii Arc acts as a power supply for the device under test, thus we had to modify the smartphone to take power from it only. We disassembled the battery separating the cells from the controller circuit. We connected the output of the Otii Arc to the controller and reconnected the circuit to the smartphone motherboard. Since the amount of required current might change quickly, we soldered 7 ceramic capacitors of 100 $\mu F$ each in parallel as close as possible to the battery controller, to avoid voltage drops. 
For this experiment, we used the Google Cloud Storage (GCS) service. We modified our application to periodically download a resource of a given size. Unlike the previous case, the amount of data sent and received is no longer symmetric. The app downloads files from three servers positioned in Italy (edge), South Carolina (cloud), and Australia (far cloud). The file size ranged from 128 to 4\,096 KB. For each file size, the execution lasted 60 minutes, and the period between two requests was equal to 20 seconds. We divided the execution time into slots of 10 minutes, computed the mean power consumption for each slot, and selected the one with the minimum value as the one least affected by external interference. In Figure~\ref{fig:google-HTTP2}, the results for the three configurations are shown. The edge is the one that achieves the lowest consumption, for almost all resource sizes, and the far cloud is the one that requires the highest power. The difference between edge and cloud is less evident, as the far cloud RTT was much higher than the other two.

\section{Discussion and Future Directions}
In this paper, we presented approaches for measuring the energy consumption of smartphones when communicating with servers in the edge-cloud continuum. We showed that in this perspective the edge is more energy-efficient than the cloud. There are some differences between the results obtained in the different scenarios. This can be due, firstly, to the specific setup configurations: communication was asymmetrical in the analytical and hardware approaches, and symmetrical in the software approach. Secondly, moving from the analytical approach to the experimental ones, we progressively introduced real-world aspects that make results not directly comparable and less predictable. The use of a real smartphone, the server implementation and configuration, the adoption of external libraries, and the actual network path all play a significant role from this point of view. We minimized some confounding factors through data hygiene and by introducing small variations in the considered evaluations. In the end, we believe that the differences between the three considered setups are indeed useful because they allow us to assess the impact of edge computing in a broader range of situations.

Our findings can be applied also to other domains, in particular the IoT domain, where battery replacement is generally more complex, and communication consumes a larger fraction of the total energy.

Future opportunities and challenges are related to the increasing presence of Artificial Intelligence (AI)~\cite{Singh2023:edge}. The case study we presented was based on relatively regular communication patterns. Additionally, the energy savings brought by reduced latency are not constant and seem to depend on the complex interaction between hardware, software, and protocol implementations. The presence of AI in the edge/cloud continuum, fed by the energy measurements provided by client devices, can help deal with the details of a multitude of slightly different systems as well as their complex interaction. This fosters the definition of new energy-saving strategies in terms of communication schemes and orchestration of resources.

\section{Acknowledgment}
Work partially supported by the Italian Ministry of University and Research (MUR) in the framework of the FoReLab project (Departments of Excellence).

\bibliographystyle{IEEEtran}
\bibliography{IEEEabrv,IC-2023-01-0006.R1_Luconi}

\begin{IEEEbiography}{Chiara Caiazza}{\,}is a research assistant at the University of Pisa, Pisa, Italy. Her research interests include network measurements and edge computing. Contact her at chiara.caiazza@ing.unipi.it. 
\end{IEEEbiography}

\begin{IEEEbiography}{Valerio Luconi}{\,}is a researcher at IIT-CNR, Pisa, Italy. His research interests include Internet measurements, Internet topology, and network monitoring. Contact him at valerio.luconi@iit.cnr.it.
\end{IEEEbiography}

\begin{IEEEbiography}{Alessio Vecchio}{\,}is an associate professor at the University of Pisa, Pisa, Italy. His research interests include network measurements and pervasive computing. Contact him at alessio.vecchio@unipi.it.
\end{IEEEbiography}

\end{document}